\documentclass[useAMS,usenatbib]{mn2e}

\usepackage{graphicx}

\title[Properties of X-ray selected BALQSOs]{Properties of X-ray selected Broad Absorption Line Quasars\thanks{Based on data from the following observatories, instruments and archives: XMM-Newton (EPIC), Subaru (Suprime-Cam), W.M. Keck Observatory (LRIS), William Herschel Telescope (WYFFOS), Hale Telescope at Palomar (WIRC), 2MASS, NASA/IPAC Infrared Science Archive; acknowledgements in full at the end.}}

\author[A. J. Blustin et al.]
  {A. J. Blustin$^{1,2}$\thanks{E-mail: ajb@ast.cam.ac.uk (AJB)}, T. Dwelly$^{3}$, M. J. Page$^{1}$, I. M. McHardy$^{3}$, N. Seymour$^{4}$,\newauthor
J. A. Kennea$^{5}$, N. S. Loaring$^{6}$, K. O. Mason$^{7,1}$, K. Sekiguchi$^{8}$ \\
  $^1$UCL Mullard Space Science Laboratory, Holmbury St. Mary, Dorking, Surrey RH5 6NT, UK\\
  $^2$Institute of Astronomy, University of Cambridge, Madingley Road, Cambridge CB3 0HA, UK\\
  $^3$Department of Physics and Astronomy, University of Southampton, Southampton SO17 1BJ, UK\\
  $^4$Spitzer Science Centre, Caltech, 1200 East California Boulevard, Pasadena, CA 91125, USA\\
  $^5$Department of Astronomy and Astrophysics, Pennsylvania State University, University Park, PA 16802, USA\\
  $^6$SALT, PO box 9, Observatory 7925, South Africa\\
  $^{7}$Science \& Technology Facilities Council, Polaris House, North Star Avenue, Swindon, Wilts SN2 1SZ, UK\\
  $^{8}$National Astronomical Observatory of Japan, Mitaka, Tokyo 181-8588, Japan\\
}
\begin{document}

\date{Accepted 2008 August 11. Received 2008 August 8; in original form 2008 June 18}

\pagerange{\pageref{firstpage}--\pageref{lastpage}} \pubyear{2008}

\maketitle

\label{firstpage}

\begin{abstract}

Broad absorption line quasars (commonly termed BALQSOs) contain the most dramatic examples of AGN-driven winds. The high absorbing columns in these winds, $\sim$~$10^{24}$~cm$^{-2}$, ensure that BALQSOs are generally X-ray faint. This high X-ray absorption means that almost all BALQSOs have been discovered through optical surveys, and so what little we know about their X-ray properties is derived from very bright optically-selected sources. A small number of X-ray selected BALQSOs (XBALQSOs) have, however, recently been found in deep X-ray survey fields. In this paper we investigate the X-ray and rest-frame UV properties of five XBALQSOs for which we have obtained XMM-Newton EPIC X-ray spectra and deep optical imaging and spectroscopy. We find that, although the XBALQSOs have an $\alpha_{ox}$ steeper by $\sim$~0.5 than normal QSOs, their median $\alpha_{ox}$ is nevertheless flatter by 0.30 than that of a comparable sample of optically selected BALQSOs (OBALQSOs). We rule out the possibility that the higher X-ray to optical flux ratio is due to intrinsic optical extinction. We find that the amount of X-ray and UV absorption due to the wind in XBALQSOs is similar, or perhaps greater than, the corresponding wind absorption in OBALQSOs, so the flatter $\alpha_{ox}$ cannot be a result of weaker wind absorption. We conclude that these XBALQSOs have intrinsically higher X-ray to optical flux ratios than the OBALQSO sample with which we compare them.

\end{abstract}


\section{Introduction}

In seeking to understand any role that AGN winds might play in the coevolution of supermassive black holes and their host galaxies, we need to study the physical properties of those winds at z~$\sim$~2 where star formation and black hole growth were at their height. It is probable that the majority of mass and energy carried in AGN winds is transported by the X-ray absorbing part of the outflow \citep[e.g.][]{blustin2007}, as the X-ray absorber has the highest column density, at least in nearby AGN, but it is usually difficult to obtain satisfactory X-ray spectra from high-redshift AGN. 

The most well-known population of distant AGN showing evidence of ionised winds is the Broad Absorption Line quasars (BALQSOs; z~$\sim$~0.1$-$6), whose winds have outflow speeds of up to 60000 km~s$^{-1}$, with line-of-sight absorbing columns of $10^{23}-10^{24}$ cm$^{-2}$ \citep[e.g.][]{krolik1999} causing the soft X-ray band to be highly absorbed \citep[see~e.g.][~and~references~therein]{gallagher2001}. The vast majority of X-ray observations of BALQSOs have been follow$-$ups after their discovery in optical surveys, and most of the \emph{XMM-Newton} and \emph{Chandra} observations of BALQSOs, with the exception of LBQS~2212-1759 \citep{clavel2006} and APM~08279+5255 \citep{chartas2002}, have been short snapshots. Our knowledge of the X-ray properties of BALQSOs is therefore based upon a small number of the very brightest optically-selected sources; the largest relevant samples to date have been published by \citet{green2001,punsly2006,gallagher2006}. 

Recently a number of X-ray selected BALQSOs (XBALQSOs) have also come to light \citep[Page~et~al.,~in~preparation;][]{barcons2003}. We use the abbreviation XBALQSOs to refer to serendipitously discovered X-ray sources whose BALQSO nature has been revealed by follow$-$up optical spectroscopic observations. It is unknown whether these represent a distinct section of the BALQSO population. Since the X-ray faintness of BALQSOs is probably due to absorption by the wind, the relative X-ray brightness of XBALQSOs could be due to their winds causing less absorption, through either having lower absorbing columns or higher ionisation parameters; XBALQSOs could also conceivably be exceptionally luminous BALQSOs with high dust extinction in the optical. There is also the possibility that their intrinsic continua simply have a higher ratio of X-ray to optical flux.

In this paper, we examine the properties of five XBALQSOs discovered in two deep X-ray survey fields: four objects in the 1$^{H}$ survey field and one in the Chandra Deep Field South (CDFS). We begin by comparing the X-ray and optical fluxes, and X-ray to optical spectral indices ($\alpha_{ox}$) of the sources to those of the largest available uniformly-analysed sample of OBALQSOs observed to date in the X-ray \citep{gallagher2006}. We then investigate whether the intrinsic extinction in the XBALQSOs differs from that in the \citet{gallagher2006} sample and from QSOs from the SDSS \citep{vandenberk2001}. In section~\ref{dynamics}, we compare the extent of intrinsic UV absorption in the XBALQSOs with that in the \citet{trump2006} sample of OBALQSOs from the SDSS. Next, in section~\ref{xray_wind_properties} we obtain estimates of the absorbing columns and ionisation parameters of the X-ray absorbing winds of the XBALQSOs, comparing the results with those of similar analyses of OBALQSOs in the literature. We then provide estimates of the mass-energy budgets of the XBALQSOs and their X-ray absorbing winds. Finally we discuss the nature of these XBALQSOs, and the implications for the population of faint BALQSOs discoverable in deep X-ray surveys.

Names, positions and basic parameters of these sources are given in Table~\ref{source_properties}. Throughout this paper, the objects are referred to by the abbreviations XBQ1-5 (X-ray selected broad absorption line quasars 1 to 5), as identified in Table~\ref{source_properties}. We use a cosmology with H$_{0}$=70~km~s$^{-1}$~Mpc$^{-1}$, $\Omega_{M}$=0.3, and $\Omega_{\Lambda}$=0.7, and all uncertainties are 1$\sigma$ unless otherwise stated.

\begin{table*}
 \centering
 \begin{minipage}{120mm}
  \caption{Basic properties of the five X-ray selected BALQSOs: IAU name and abbreviated name for use in this paper; RA; Dec; r, offset between the optical and X-ray positions in arcseconds; redshift (see section~\ref{optical_data}); Galactic extinction using reference pixel in \citet{schlegel1998}; de-reddened AB $B$ and $K_s$ magnitudes, where the figures for XBQ5 are from the MUSYC survey (\citealt{gawiser2006}; Taylor et al., in preparation; see section~\ref{optical_data}); C$_{x}$, number of background-subtracted X-ray counts in the 0.2$-$12~keV range.}
  \begin{tabular}{@{}llllll@{}}
  \hline
  Name  & & & RA & Dec & r  \\
 \hline
CXOU J014546.9-043031    & & XBQ1 & 01:45:46.88 & -04:30:30.50 & 0.50  \\
CXOU J014517.1-042503    & & XBQ2 & 01:45:17.05 & -04:25:03.20 & 0.91  \\
CXOU J014553.9-043726    & & XBQ3 & 01:45:53.93 & -04:37:26.20 & 0.74   \\
CXOU J014546.5-042450    & & XBQ4 & 01:45:46.54 & -04:24:49.50 & 0.78   \\
CXOCDFS J033209.5-274807 & & XBQ5 & 03:32:09.46 & -27:48:06.80 & 0.28\footnote{Table 2, \citet{giacconi2002}} \\
\hline
  \hline
  Name & z & E(B-V) & $B$ & $K_s$ & C$_{x}$\\
 \hline
 XBQ1 & 2.63  & 0.023 & 21.8 $\pm$ 0.1 & 21.1 $\pm$ 0.2 & 208  \\
 XBQ2 & 1.793 & 0.023 & 21.4 $\pm$ 0.1 & 20.7 $\pm$ 0.2 & 154  \\
 XBQ3 & 1.40  & 0.022 & 21.5 $\pm$ 0.1 & 19.6 $\pm$ 0.2 & 74  \\
 XBQ4 & 2.64  & 0.024 & 20.1 $\pm$ 0.1 & 19.5 $\pm$ 0.2 & 137  \\
 XBQ5 & 2.82  & 0.009 & 21.33 $\pm$ 0.05 & 20.00 $\pm$ 0.06 & 1115  \\
\hline
\end{tabular}
\end{minipage}
\label{source_properties}
\end{table*}

\section{Observations and data reduction}

The 1$^{H}$ field is centred at 01$^{H}$45$^{m}$27$^{s}$ -04$^{\circ}$34$\arcmin$42$\arcsec$; EPIC data from this field were obtained as part of \emph{XMM-Newton} Guaranteed Time observations as a `blank field' X-ray survey. A mosaic of four 30~ks Chandra datasets was also obtained for the field. These data were primarily used to determine sub-arcsecond precision positions for the \emph{XMM-Newton} sources, whose astrometry could be tied to that of the Chandra sources, and in turn to sources detected in deep optical imaging. Follow-up optical spectroscopic observations of over 100 X-ray sources in the 1$^{H}$ field have revealed four BALQSOs; in addition, one BALQSO has been discovered in follow-up spectroscopy of Chandra sources in the CDFS (out of 168 objects with spectroscopic identifications; \citealt{szokoly2004}). The \emph{XMM-Newton} pointings of the CDFS were centred at 03$^{H}$32$^{m}$27$^{s}$ $-$27$^{\circ}$48$\arcmin$55$\arcsec$, and we obtained these data from the XSA\footnote{XMM-Newton Science Archive, http://xmm.esac.esa.int/xsa/}. 

\begin{table*}
 \centering
 \begin{minipage}{100mm}
  \caption{\emph{XMM-Newton} observations of the 1$^{H}$ and CDFS fields; the amount of good time is listed for each EPIC instrument.}
  \begin{tabular}{@{}llllll@{}}
  \hline
Field & Observation & Date & MOS1 & MOS2 & pn \\
  &   &   & Exp (ks) & Exp (ks) & Exp (ks) \\
\hline
1$^{H}$  & 0109661101 & 2002-12-25 & 47.7 & 47.8 & 42.0  \\
1$^{H}$  & 0109661201 & 2002-07-16 & 47.1 & 47.5 & 41.8  \\
1$^{H}$  & 0109661401 & 2002-07-17 & 3.0  & 3.3  & 2.7   \\ 
1$^{H}$  & Total good time &            & 97.8 & 98.6 & 86.6  \\
\hline
CDFS & 0108060401 & 2001-07-27 & 20.2  & 20.2  & 13.4 \\
CDFS & 0108060501 & 2001-07-28 & 39.6  & 40.7  & 31.7 \\
CDFS & 0108060601 & 2002-01-13 & 47.6  & 47.6  & 42.0 \\
CDFS & 0108060701 & 2002-01-14 & 73.5  & 73.1  & 67.9 \\
CDFS & 0108061801 & 2002-01-16 & 54.8  & 54.8  & 36.5 \\
CDFS & 0108061901 & 2002-01-17 & 42.6  & 42.6  & 38.9 \\
CDFS & 0108062101 & 2002-01-20 & 43.2  & 43.5  & 40.6 \\
CDFS & 0108062301 & 2002-01-23 & 74.3  & 72.8  & 69.3 \\
CDFS & Total good time &     &       395.8 & 395.2 & 340.2 \\
\end{tabular}
\end{minipage}
\label{obs_details}
\end{table*}

\subsection{\emph{XMM-Newton} EPIC spectral extraction}

The X-ray data reduction and spectral extraction followed the same process as was used for the 13$^{H}$ \emph{XMM-Newton} deep survey field \citep{mchardy2003,loaring2005,page2006}. The raw event lists were processed using the \emph{EMCHAIN} and \emph{EPCHAIN} tasks under SAS V6.0; data affected by high background were removed by filtering out time intervals with count-rates above 5~keV of greater than 2.2~photons~s$^{-1}$ and 4~photons~s$^{-1}$ for MOS and pn respectively. Although $\sim$~200~ks of \emph{XMM-Newton} data are available for the 1$^{H}$ field, about half of that is affected by high background; for the CDFS, about 75\% of the original $\sim$~500~ks of exposure time is usable. The amount of good time for each observation and EPIC instrument is listed in Table~\ref{obs_details}. The X-ray source lists were generated using an iterative procedure that models the EPIC background and detects sources; for details, see \citet{loaring2005}. Full details of the \emph{XMM-Newton} data reduction for the CDFS are given by \citet{dwellypage2006}.  

In extracting MOS spectra, all valid events (PATTERN = 0$-$12) were selected, whereas for the pn, single and double events (PATTERN = 0$-$4) were used for channel energies greater than 0.4 keV, and single events only (PATTERN = 0) for energies below this. After the generation of response matrices and effective area files using \emph{RMFGEN} and \emph{ARFGEN} respectively, the spectra from each instrument for each source were combined using the method of \citet{page2003}. The co-added 0.2$-$12~keV spectra were grouped to have at least 20 counts per bin, except in the case of XBQ3 where we decreased this to 18 counts per bin in order to have sufficient points to fit our absorber model (section~\ref{xray_wind_properties}).

\subsection{Optical observations and data reduction}
\label{optical_data}

For the four sources in the 1$^{H}$ field, $B$ magnitudes were calculated from Subaru Suprime-Cam \citep{miyazaki2002} imaging, and $K_s$ magnitudes were derived from Palomar 200\arcsec\, WIRC \citep{wilson2003} data. The $B$-band data were collected in October/November 2002, and the $K_s$ data were taken in September 2005. We list the optical magnitudes that we obtained for XBQ1$-$4 in Table~\ref{source_properties}, alongside magnitudes for XBQ5 obtained from the MUSYC survey (\citealt{gawiser2006}; Taylor et al., in preparation)\footnote{http://www.astro.yale.edu/MUSYC/}. The photometry for XBQ1$-$4 used 3$\arcsec$ diameter apertures, and the magnitudes were converted from the Vega to the AB systems using the appropriate transformations from \citet{fukugita1996}, with the exception of $K_s$ where we calculate $K_s$(Vega) = $K_s$(AB) $-$ 1.86 using the $K_s$ pass-band information and detector response given on the Palomar website\footnote{http://www.astro.caltech.edu/palomar/}. For our (relatively bright) BALQSOs, the errors on the magnitudes are dominated by systematics, primarily uncertainty on zeropoints. For XBQ5, to account for calibration uncertainties for this MUSYC field \citep[see~e.g.][]{hildebrandt2006}, the magnitude errors incorporate an additional 0.05 mag error added in quadrature to the nominal photometric uncertainties. The XBALQSO magnitudes were corrected for Galactic reddening using the dust maps of \citet{schlegel1998}\footnote{http://irsa.ipac.caltech.edu/applications/DUST/}. The magnitudes are not host-subtracted; we assume that the host contribution is negligible, since the XBALQSOs are all point-like sources, and there is no evidence of host galaxy light in the spectra.

\begin{figure*}
\includegraphics[width=100mm,angle=-90]{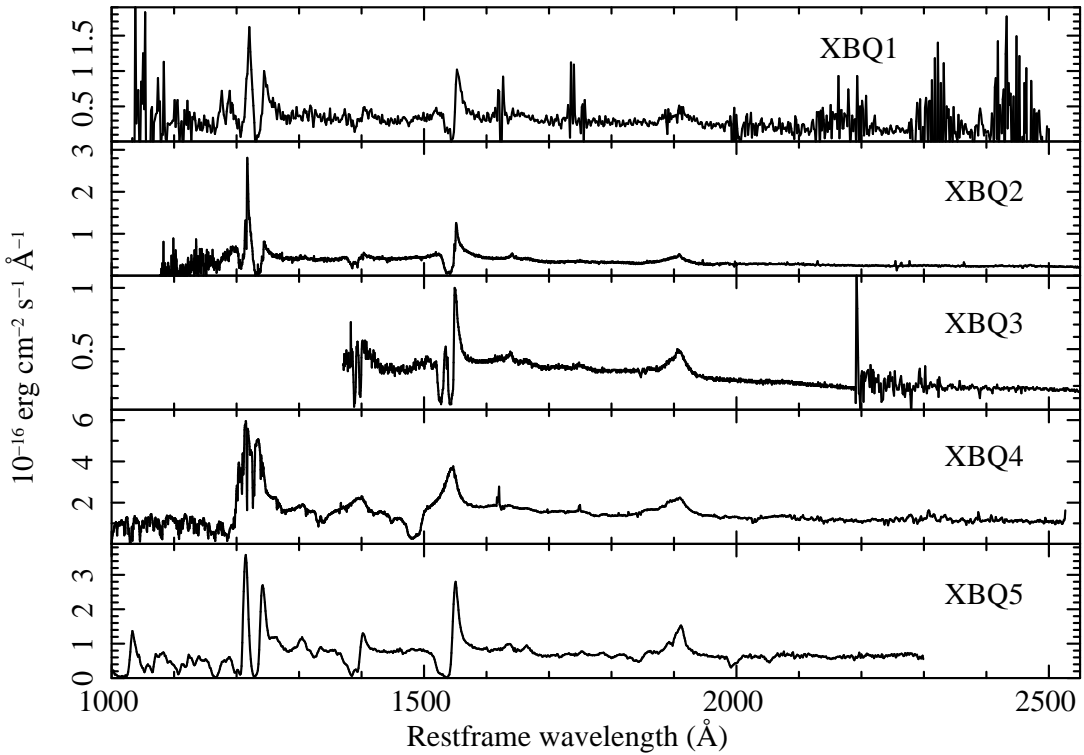}
 \caption{Plots of the rest-frame UV spectra of the five XBALQSOs, not corrected for Galactic reddening.}
\label{all_restframe}
\end{figure*}

Observations using several different multi-object spectrometers on 4~m and 10~m class telesopes have been carried out as part of programme to classify and obtain redshifts for the counterparts to X-ray and radio sources in the 1$^{H}$ field. An optical spectrum of XBQ1 was obtained from WHT-AF2/WYFFOS on 2003 November 25. The observation used the 316R grating, covering the 3600-9000~\AA\ range with a resolution of 8~\AA, with a total exposure time of 4900~s. For XBQ2, XBQ3 and XBQ4 the data came from the Keck Low Resolution Imaging Spectrograph (LRIS) operated in Multi Object Spectroscopy (MOS) mode, with spectra taken on 2003 October 29 and 30, with 7200~s exposure per mask. The $\sim$~3500$-$9500~\AA\, range is covered by the combination of blue and red arms, with a spectral resolution of $\sim$~5~\AA\, in the blue arm and $\sim$~9~\AA\ in the red arm. Data reduction was performed with IRAF \citep{tody1993} using the NOAO \emph{twodspec apextract} and the LRIS-specific \emph{wmkolris} packages. Full details of the Keck-LRIS observations and data reduction are given by \citet{dwelly2006}. We obtained an optical spectrum of XBQ5 from the \citet{szokoly2004} catalogue of CDFS optical spectra\footnote{http://www.mpe.mpg.de/CDFS/data/62.html}; those data were obtained using VLT-FORS \citep{appenzeller1998}, with an exposure time of 10800~s, a spectral coverage of $\sim$~3600$-$9000~\AA\, and a spectral resolution of $\sim$~20~\AA\, at 5600~\AA. All spectra were normalised according to their measured $B$-magnitudes (Table~\ref{source_properties}). We measured the systemic redshifts of XBQ1$-$5 from their optical spectra, which are shown plotted in the restframe in Fig.~\ref{all_restframe}, and are listed in Table~\ref{source_properties}; our redshift for XBQ5 is close to that quoted by \citet{szokoly2004}; the values are 2.82 and 2.81 respectively.

\section{Observed optical fluxes, X-ray fluxes and $\alpha_{ox}$}
\label{fluxes_alpha_ox}

We derived restframe 2500~\AA\ flux densities from the optical spectra (from the data where possible, or by extrapolation of a continuum fit where necessary), and corrected them for Galactic reddening as in section~\ref{optical_data}. In order to generate $\alpha_{ox}$ indices which were directly comparable with those listed by \citet{gallagher2006} for their sample of OBALQSOs, we estimated the rest-frame 2~keV X-ray flux density for our XBALQSOs using a method analogous to theirs. Accordingly, we fitted a Galactic absorbed power-law to the observed frame 0.5$-$8~keV X-ray spectra of each source, using SPEX 2.00.11 \citep{kaastra1996}\footnote{http://www.sron.nl/divisions/hea/spex/}, and used this power-law to calculated the unabsorbed 2~keV X-ray flux densities. We then calculated $\alpha_{ox}$ \citep{tananbaum1979}; the resulting values are listed in Table~\ref{alpha_ox}, along with the observed frame unabsorbed 0.5$-$8~keV X-ray fluxes.

\begin{table*}
 \centering
 \begin{minipage}{150mm}
  \caption{Observed X-ray fluxes, X-ray and UV flux densities, and $\alpha_{ox}$ for the XBALQSOs, derived from the simple X-ray power-law fit and UV spectral measurements described in section~\ref{fluxes_alpha_ox}: Name; N$_{H, Gal}$, Galactic equivalent Hydrogen column (using {\it nh} FTOOL with the LAB Survey of Galactic HI map; \citealt{kalberla2005}) in 10$^{20}$ cm$^{-2}$; $\Gamma_{0.5-8~keV}$, spectral index of power-law fitted to observed frame 0.5$-$8~keV; Log $F_{X}$(0.5-8~keV), 0.5$-$8~keV flux of Galactic absorbed power-law fit in erg cm$^{-2}$ s$^{-1}$; Log F$_{\nu}$(2500~\AA), energy flux density at 2500~\AA\, in erg cm$^{-2}$ s$^{-1}$ Hz$^{-1}$; Log F$_{\nu}$(2~keV), energy flux density at 2~keV in erg cm$^{-2}$ s$^{-1}$ Hz$^{-1}$; $\alpha_{ox}$.}
  \begin{tabular}{@{}lllllll@{}}
  \hline
Name & N$_{H, Gal}$ & $\Gamma_{0.5-8~keV}$ & Log F$_{X}$(0.5-8~keV) & Log F$_{\nu}$(2500~\AA) & Log F$_{\nu}$(2~keV) & $\alpha_{ox}$  \\
 \hline
XBQ1 & 2.28  & 1.5 $\pm$ 0.2      & -14.43$^{+0.06}_{-0.08}$ & -28.5  & -32.9 $\pm$ 0.1     & -1.69   \\
XBQ2 & 2.38  & 0.8 $\pm$ 0.4      & -14.3$^{+0.1}_{-0.2}$    & -28.3  & -33.2 $\pm$ 0.2     & -1.88   \\
XBQ3 & 2.18  & 0.5$^{+0.4}_{-0.5}$  & -14.4$^{+0.1}_{-0.2}$   & -28.4  & -33.4$^{+0.2}_{-0.3}$ & -1.90  \\
XBQ4 & 2.45  & 1.4 $\pm$ 0.4      & -14.5 $\pm$ 0.1        & -27.6  & -33.0 $\pm$ 0.2     & -2.07   \\
XBQ5 & 0.698 & 1.18 $\pm$ 0.06    & -14.18 $\pm$ 0.03      & -27.9  & -32.9 $\pm$ 0.04    & -1.90  \\
\hline
\end{tabular}
\end{minipage}
\label{alpha_ox}
\end{table*}

\begin{figure}
\includegraphics[width=55mm,angle=-90]{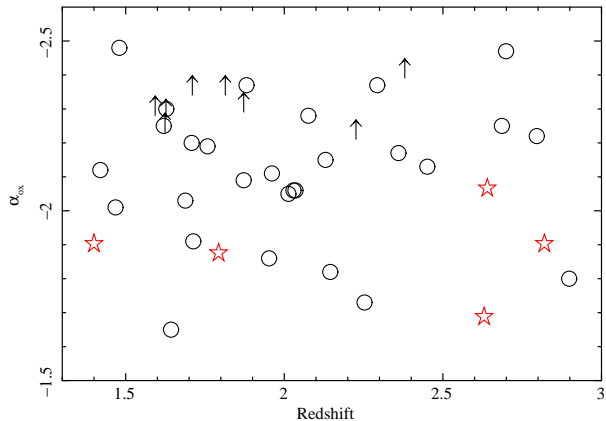}
 \caption{A comparison of $\alpha_{ox}$ of the XBALQSOs (red stars) with the \citet{gallagher2006} sample of OBALQSOs (black circles; black arrows are upper limits), plotted versus redshift.}
\label{opt_xray_comp}
\end{figure}

Fig.~\ref{opt_xray_comp} shows the XBALQSO $\alpha_{ox}$ values compared with those for the \citet{gallagher2006} sample of OBALQSOs. The KS-test probability that $\alpha_{ox}$ for the XBALQSOs and OBALQSOs are drawn from the same population is 0.053, implying marginal consistency. The average $\alpha_{ox}$ of a typical QSO is $\sim$~-1.4 \citep[see~e.g.][]{marconi2004}; the median $\alpha_{ox,XBALQSO}$ is -1.90 (standard deviation 0.13), somewhat flatter than the median $\alpha_{ox,OBALQSO}$ which is -2.20 (standard deviation 0.21). This difference of 0.3 in $\alpha_{ox}$ between the two samples of BALQSOs implies that, for a given $F_{\nu}$(2500), one gets $\sim$~6 times more X-ray flux from an average XBALQSO than from an average OBALQSO.

\section{Intrinsic extinction properties in the XBALQSOs}
\label{extinction_properties}

We investigated whether or not the XBALQSOs had excess intrinsic extinction by calculating their (dereddened with respect to Galactic extinction) $B-K_s$ colours and comparing them with those of the \citet{gallagher2006} sample of OBALQSOs, and more generally of QSOs in the SDSS \citep{vandenberk2001}. The XBALQSO $B$ and $K_s$ magnitudes (from Table~\ref{source_properties}) are both in the AB system. 

We adopt the UK Schmidt $B_j$ magnitudes for the OBALQSOs taken from \citet{gallagher2006} (in turn taken from \citealt{hewett1995}). The blue cutoffs of the $B_j$ and Johnson $B$ bandpasses are similar \citep{couchnewell1980}, but the $B_j$ system extends $\sim$~500~\AA\ further into the red. We estimate that $B-B_j<0.1$ for QSOs in the redshift range of interest ($\sim1.3-3.0$). $K_s$ magnitudes for the OBALQSOs were obtained from the 2MASS point source catalogue \citep{skrutskie2006}; values were available for 29 out of 36 sources in the sample. These were converted to AB using $K_s$(AB) = $K_s$(Vega) + 1.85 \citep{blanton2005}. The magnitudes were dereddened using reference pixel E(B-V) values from \citet{schlegel1998}.

\begin{figure}
\includegraphics[width=55mm,angle=-90]{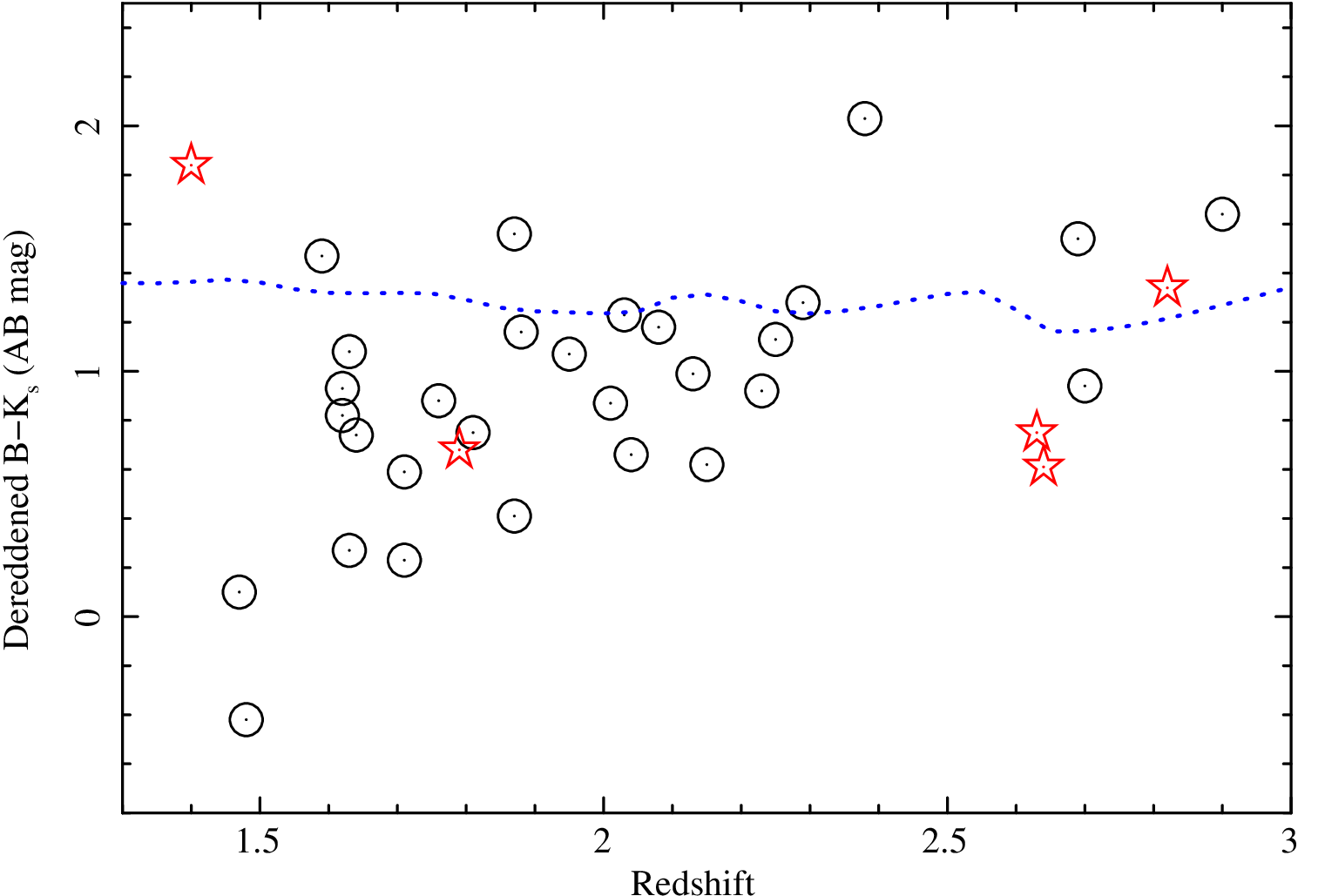}
 \caption{A comparison of the dereddened (with respect to Galactic extinction) $B-K_s$ colours of our XBALQSOs (red stars) with those of the \citet{gallagher2006} sample of OBALQSOs (black circles) with available $K_s$ magnitudes, plotted versus redshift. The median $B-K_s$ colour of SDSS QSOs is overplotted (blue dotted line; \citealt{vandenberk2001}).}
\label{extinction_comp}
\end{figure}

The $B-K_s$ colours for the XBALQSOs and OBALQSOs are plotted in Fig.~\ref{extinction_comp}. The KS-test probability of the XBALQSO colours and \citet{gallagher2006} OBALQSO colours being drawn from the same distribution is p$_{KS}$ = 0.9, implying consistency.

\section{Dynamics of the UV BAL, and properties of the UV emission lines}
\label{dynamics}

\begin{figure*}
\includegraphics[width=170mm,angle=-90]{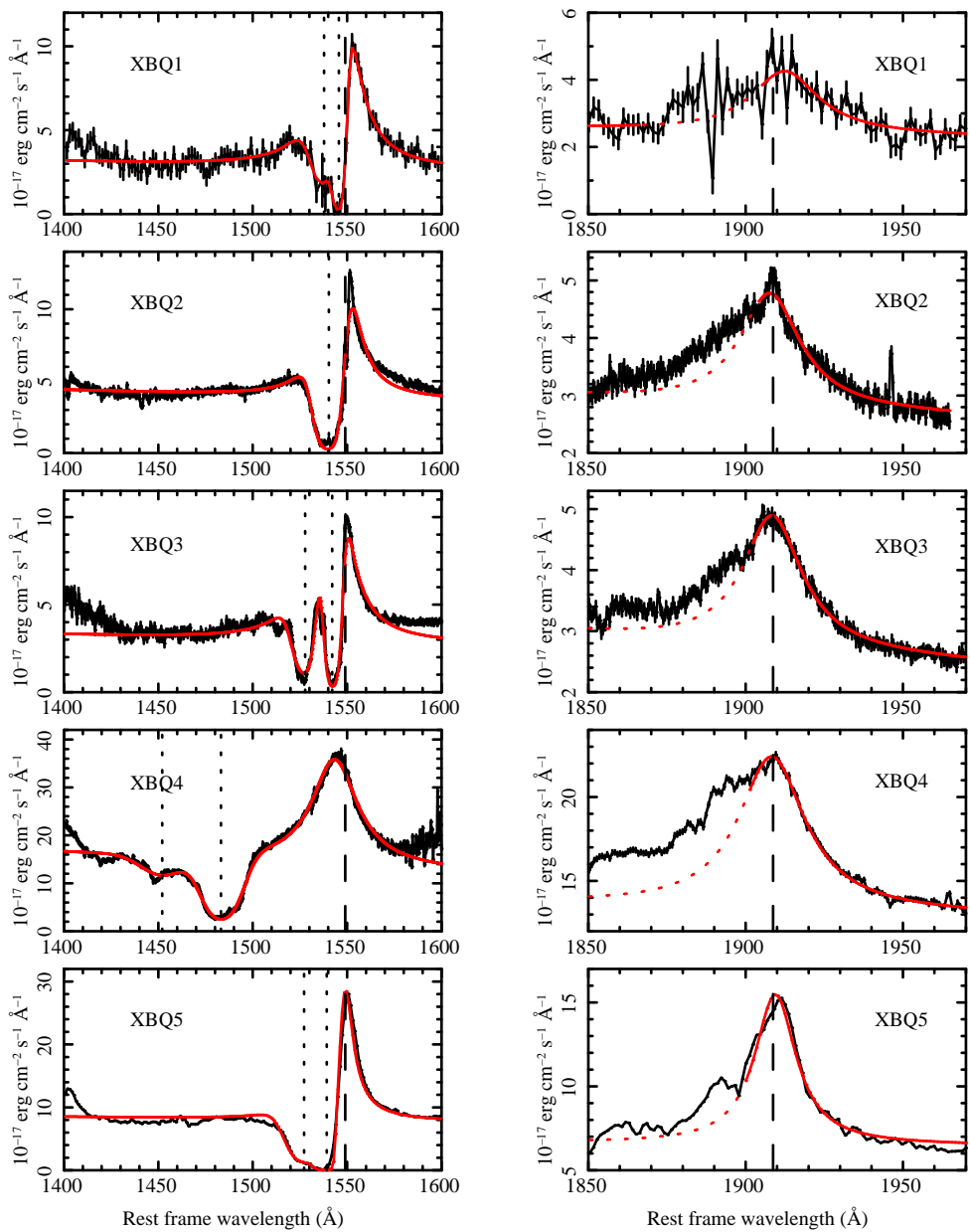}
 \caption{Plots of the C~IV $\lambda\lambda$1549 (left panels) and the C~III] $\lambda\lambda$1909 (right panels) line regions of the five XBALQSOs, in the source rest-frames, with fitted models superimposed (red). The vertical dashed lines indicate the lab wavelengths for C~IV $\lambda\lambda$1549 and C~III] $\lambda\lambda$1909 respectively, and the vertical dotted lines indicate the fitted centroids of the absorption features. The dotted sections of the models in the right panels denote regions excluded from the emission line fit in order to avoid contamination from the Si III] $\lambda\lambda$1892 and Al III $\lambda\lambda$1857 emission lines.}
\label{opt_fit_spectra}
\end{figure*}

We obtained the line-of-sight velocities and velocity widths of the principal outflow components in each XBALQSO from the optical (rest-frame UV) spectra. We used the \emph{specfit} \citep{krissspecfit} routine in IRAF to fit the spectral region around the C~IV $\lambda\lambda$1549 emission line, in each source, with a model incorporating Galactic E(B-V), a power-law continuum, a Lorentzian emission line, and Gaussian absorption line components for the absorber; all sources except for XBQ2 have two principal outflowing components. The resulting fitted absorption properties are listed in Table~\ref{uv_abs_lines}, and the fitted models are overplotted on the rest-frame spectra in Fig.~\ref{opt_fit_spectra}.

\begin{table*}
 \centering
 \begin{minipage}{160mm}
  \caption{Intrinsic UV absorption feature properties, where the fitted features are approximated as single broadened gaussians: source name; number identifying absorption phase; $\lambda_{meas}$, observed frame wavelength in \AA; rest frame FWHM in km s$^{-1}$; $\tau$, optical depth of line; $v_{shift}$, velocity shift of line in km s$^{-1}$;  BI, Balnicity Index \citep{weymann1991} in km s$^{-1}$; AI, Absorption Index \citep{trump2006} in km s$^{-1}$ (see section~\ref{BI_AI} for details of the calculation of the BI and AI and their uncertainties). }
\label{uv_abs_lines}
  \begin{tabular}{@{}llllllll@{}}
  \hline
  Name  & Absorption phase & $\lambda_{meas}$  & FWHM           & $\tau$          & $v_{shift}$       & BI            &            AI  \\
 \hline
XBQ1    & 1                & 5610.4 $\pm$ 0.9 & 1140 $\pm$ 100 & 3.3 $\pm$ 0.4   & -670 $\pm$ 50    & 366 $\pm$ 78  & 3305 $\pm$ 132 \\
        & 2                & 5582   $\pm$ 3   & 2610 $\pm$ 190 & 1.6 $\pm$ 0.2   & -2170 $\pm$ 150  &               &                \\
XBQ2    & 1                & 4302.1 $\pm$ 0.2 & 2430 $\pm$ 30  & 3.9 $\pm$ 0.1   & -1680 $\pm$ 20   & 701 $\pm$ 8   & 3483 $\pm$ 9  \\
XBQ3    & 1                & 3666.8 $\pm$ 0.4 & 2390 $\pm$ 80  & 1.97 $\pm$ 0.05 & -4100 $\pm$ 30   & 2143 $\pm$ 18 & 4799 $\pm$ 17  \\
        & 2                & 3701.1 $\pm$ 0.2 & 1460 $\pm$ 30  & 4.1 $\pm$ 0.1   & -1330 $\pm$ 20   &               &               \\
XBQ4    & 1                & 5398.8 $\pm$ 0.5 & 3790 $\pm$ 60  & 1.86 $\pm$ 0.04 & -12740 $\pm$ 30  & 3631 $\pm$ 11 & 5144 $\pm$ 11  \\
        & 2                & 5286 $\pm$ 2     & 5210 $\pm$ 370 & 0.31 $\pm$ 0.01 & -18750 $\pm$ 100 &               &               \\
XBQ5    & 1                & 5880 $\pm$ 1     & 1940 $\pm$ 130 & 7.5 $\pm$ 0.6   & -1900 $\pm$ 70   & 3011\footnote{Uncertainties on BI or AI cannot be calculated for this source as the archival optical spectral points \citep{szokoly2004} were not supplied with error bars.}         & 6223 \\
        & 2                & 5834 $\pm$ 8     & 3220 $\pm$ 700 & 2.2 $\pm$ 0.2   & -4220 $\pm$ 390  &               &               \\
\hline
\end{tabular}
\end{minipage}
\end{table*}

\begin{table*}
 \centering
 \begin{minipage}{160mm}
  \caption{Spectral regions for the UV spectral fits, and FWHMs of the UV emission lines, rounded to the nearest 10 km~s$^{-1}$: Object name; Fit ranges for C~IV, observed frame spectral ranges for the C~IV $\lambda\lambda$1549 region fit; Fit ranges for {C~III]}, observed frame spectral ranges for the C~III] $\lambda\lambda$1909 region fit; FWHM$_{C~III]}$, fitted FWHM of C~III] $\lambda\lambda$1909 in km~s$^{-1}$; FWHM$_{C~IV,~meas}$, fitted FWHM of C~IV $\lambda\lambda$1549 in km~s$^{-1}$; FWHM$_{C~IV,~pred}$, FWHM of C~IV $\lambda\lambda$1549, in km~s$^{-1}$, predicted according to the finding of \citet{brotherton1994} that FWHM$_{C~III]}$ = 1.2 $\times$ FWHM$_{C~IV}$.}
\label{lines}
  \begin{tabular}{@{}llllll@{}}
  \hline
Name & Fit ranges for C~IV & Fit ranges for C~III] & FWHM$_{C~III]}$ & FWHM$_{C~IV,~meas}$ & FWHM$_{C~IV,~pred}$  \\
 \hline
XBQ1 & 5185$-$5783, 7386$-$7662, & 6407$-$6774, 6910$-$7172  & 4060 & 3870 & 3380 \\
     & 8068$-$8241               &   & & & \\
XBQ2 & 4024$-$4157, 4180$-$4367, & 4952$-$5158, 5319$-$5466  & 3640 & 2600 & 3040 \\
     & 4932$-$5098               &   & & &   \\
XBQ3 & 3438$-$3775, 5067$-$5236  & 4278$-$4367, 4564$-$4885  & 3590 & 2930 & 2990 \\
XBQ4 & 4740$-$4758, 5132$-$5698, & 6505$-$6626, 6918$-$9143  & 4050 & 5750 & 3380 \\
     & 5855$-$5905               &   & & &   \\
XBQ5 & 4784$-$4822, 5456$-$5529, & 6577$-$6830, 7256$-$7563, & 2500 & 730  & 2090 \\
     & 5636$-$5971, 8065$-$8700  & 7973$-$8628               & & &  \\
\hline
\end{tabular}
\end{minipage}
\end{table*}

The width of the C~IV emission line can be used to estimate the black hole mass, and thus the Eddington luminosity. Since the fitted width of C~IV is likely to be affected by the presence of the absorption features, we also obtained the width of CIII] $\lambda\lambda$1909 in each case, and used these, with the result of \citet{brotherton1994} that, in intermediate-redshift (0.9$<$z$<$2.2) quasars, FWHM$_{CIII]}$ = 1.2 $\times$ FWHM$_{CIV}$, to provide a predicted C~IV FWHM in each case. The C~III] fits used the same basic model as the C~IV fits, minus the intrinsic absorption components and with a locally-fitted power-law continuum. Much of the blue wing of each C~III] line was omitted from the fit in each case, in order to exclude contamination from the Si III] $\lambda\lambda$1892 and Al III $\lambda\lambda$1857 emission lines. 

The fitted and predicted FWHMs, alongside the spectral ranges used for the emission and absorption line fits, are listed in Table~\ref{lines}. The differences between them are probably partly due to the uncertainty introduced by the presence of absorption features; although the fits should take account of the extent of absorption in fitting the emission lines, the contribution of each to the spectral shape becomes more ambiguous as the absorption encroaches further towards the emission line centroid. Intrinsic scatter around the \citet{brotherton1994} relation is likely to be another reason, and probably also the fact that three of our XBALQSOs are outside the redshift range for which the relation was derived.

\subsection{How does the amount of intrinsic UV absorption compare with that in the OBALQSO population?}
\label{BI_AI}

In order to compare the extent of rest-frame UV absorption in the five XBALQSOs with that in the population of OBALQSOs, we calculated the Balnicity Index \citep[BI;][]{weymann1991} and Absorption Index \citep[AI;][]{trump2006} for each object. The BI and AI are measures of the equivalent width of the BAL absorption; the BI is defined to include features at blueshifts between 3000 and 25000 km~s$^{-1}$, and only that part of a trough beyond 2000 km~s$^{-1}$ in width; the more recently-defined AI measure includes all absorption features wider than 1000 km~s$^{-1}$ and between zero and 29000 km~s$^{-1}$ blueshift. 

For each object, we used the model fits described in Section~\ref{dynamics} to represent the local continua for the purposes of measuring the BI and AI. We estimated the uncertainties on BI and AI according to Equations (5) and (7) in \citet{trump2006} respectively. In practice, uncertainties introduced by the continuum placement are likely to be greater sources of error than the statistical uncertainties.

There are difficulties with both the BI and AI as defining criteria for the BALQSO population \citep[e.g.][]{knigge2008}, but for our purposes, the BIs and AIs listed by \citet{trump2006} provide a useful comparison with a very large consistently-defined sample.

\begin{figure}
\includegraphics[width=58mm,angle=-90]{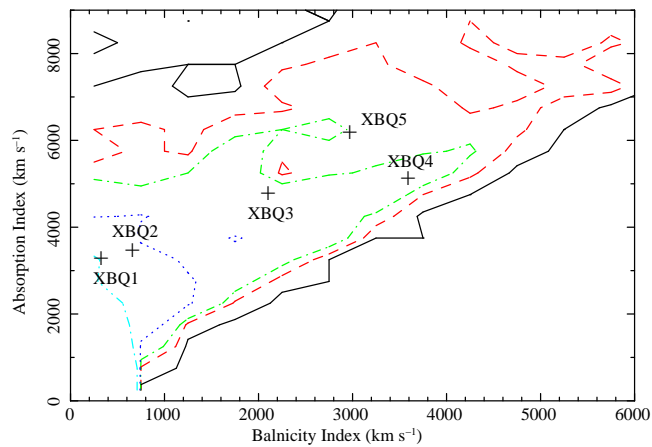}
 \caption{Two-dimensional histogram, in 500 km~s$^{-1}$ width bins, of BI versus AI for 4843 C~IV BALS in the SDSS catalogue \citep{trump2006}, with the positions of the five X-ray selected BALQSOs in our sample marked. The contours are 1 (black; solid line), 5 (red; dashed line), 10 (green; dot$-$dashed line), 25 (dark blue; dotted line), and 100 (light blue; triple$-$dot$-$dashed line) sources per BI$-$AI bin. The plotted ranges in BI and AI encompass $\sim$~98\% of the \citet{trump2006} sample.}
\label{BI_AI_plot}
\end{figure}

The resulting values for BI and AI of each source are listed in Table~\ref{uv_abs_lines}, and in Fig.~\ref{BI_AI_plot} they are overplotted on the BI-AI distribution of 4843 C~IV BALS in the SDSS catalogue \citep[from][]{trump2006}\footnote{The J/ApJS/165/1/table4 catalogue in Vizier: http://vizier.u-strasbg.fr/viz-bin/VizieR}. Our sources all fall within the BI-AI distribution of the optically-selected BALQSO population. Using a 2D Kolmogorov-Smirnov (KS) test \citep{fasano1987}, the probability that the XBALQSOs and the whole set of 4843 SDSS BALQSOs are drawn from the same population is p$_{2DKS}$ = 0.009; comparing the XBALQSOs with SDSS objects with BI $>$ 0, we obtain p$_{2DKS}$ = 0.3. The XBALQSOs are thus statistically consistent, in terms of balnicity at least, with the subset of BALQSOs from \citet{trump2006} that would be classed as `classical' BALQSOs.

\section{The properties of the X-ray absorbing wind}
\label{xray_wind_properties}

Due to the small numbers of X-ray counts in our XBALQSOs, it is of course possible to obtain good $\chi^{2}_{red}$ with simple power-law fits to the data. Our focus, though, is on constraining the properties of the ionised wind, rather than comparing the goodness of fit of different possible models. We believe that this approach is justifiable for the following reasons: from the UV spectra (and indeed by definition as BALQSOs), we know that the XBALQSOs have outflowing, ionized winds with high optical depths; in the case of nearby AGN, the UV-absorbing winds are always accompanied by X-ray absorption \citep{crenshaw1999}, and this is likely to be the case in BALQSOs as well \citep[e.g.][]{gallagher2006}; our photoionized absorber models essentially already cover the possibility of neutral absorption, in the limit of very low ionization. Given this information, presenting separate fits of simpler models would seem redundant.

If we are therefore going to fit ionised wind models, we will need to make certain assumptions due to the low signal-to-noise of the data. We cannot use X-ray absorption features to measure the dynamical properties of the wind, or to study the wind's ionisation structure in as much detail as is possible for nearby AGN \citep[see~e.g.][]{netzer2003}. Our procedure for modelling the wind X-ray absorption, and the assumptions that we make, are as follows.

\begin{enumerate}
  \item We assume that the X-ray absorber has the same velocity structure as the UV absorber (as measured in Section~\ref{dynamics}), by analogy with nearby AGN where the dynamics of the UV-absorbing gas are often similar to the situation in the X-ray absorber \citep[see~e.g.][]{kraemer2002,kraemer2003,gabel2003,blustin2007};
  \item We calculate photoionised absorber models using XSTAR \citep{kallman1982}, assuming canonical Spectral Energy Distributions (SEDs; in this case those of \citealt{marconi2004}; Section~\ref{seds})); 
  \item We fit the photoionised absorber models to the X-ray spectra using SPEX 2.00.11 \citep{kaastra1996}, assuming that the X-ray outflow phases have the same line-of-sight velocities and velocity widths as the UV BAL components, and using the canonical SEDs as renormalizable continuum models (Section~\ref{wind_fitting}).
  \item Since we have no independent information about the range of column densities and ionisation parameters in the X-ray absorbing wind, we also assume in the fitting that all dynamical phases in a given source have the same column density and ionisation parameter. This is probably the weakest assumption that we make, since, at least in nearby AGN, there are usually two or more ionisation regimes in the wind \citep[e.g.][]{holczer2007,blustin2005}. Fitting a single ionisation phase model is likely to give an absorber with an intermediate ionization parameter and column; the accuracy and implications of this assumption are issues that we intend to address in future work. For the purposes of this analysis, the single zone model gives us a convenient comparison with the models fitted to OBALQSOs in the literature.
\end{enumerate}

The stages of the analysis are described in full detail in the following sections.

\subsection{Modelling the ionising continuum}
\label{seds}

The intrinsic SEDs of the five sources were modelled on the canonical AGN SED described by \citet{marconi2004}. The UV/optical range was represented with a series of power-law segments; between 1 $\mu$m and 1250~\AA\ the power-law index $\alpha$ was -0.44 (where F$_{\nu}\propto\nu^{\alpha}$), and between 1250~\AA\ and 500~\AA\ the index was -1.76. Longward of 1 $\mu$m the spectrum has an index of 2, to represent the Rayleigh-Jeans tail. The normalisation of the UV/optical range was provided by the dereddened (rest-frame) 2500~\AA\ flux densities from the optical spectra.

The X-ray SED between 2~keV and 100~MeV was created from a SPEX 2.00.11 continuum model. This consisted of a $\Gamma$=1.9 power-law plus a reflection component, generated with the SPEX {\it refl} model, from neutral material at an inclination of cos $i$ = 0.5 and a reflection fraction of 1. The continuum is cut off exponentially at 500~keV.

The UV/optical and X-ray SEDs were joined by a power-law between 500~\AA\ and 2 keV. The normalisation of the X-ray band relative to the optical was set using an intrinsic $\alpha_{ox}$ index; this was estimated from the dereddened 2500~\AA\ (rest-frame) luminosity $L_{\nu}$(2500~\AA) according to the expression given by \citet{vignali2003}:

\begin{equation}
  \alpha_{ox} = -0.11 {\mathrm {log}} L_{\nu}(2500~{\mathrm {\AA}}) + 1.85
\end{equation}

Given that the X-ray to optical ratios of these sources are higher than expected, it is worth asking whether we could use our observed $\alpha_{ox}$ indices to determine the relative normalisation of the X-ray and optical parts of the SED. This is unfortunately not feasible, since the presence of X-ray absorption in the low signal-to-noise spectra prevents us from unambiguously determining the intrinsic level of the X-ray continuum.

\begin{figure}
\includegraphics[width=60mm,angle=-90]{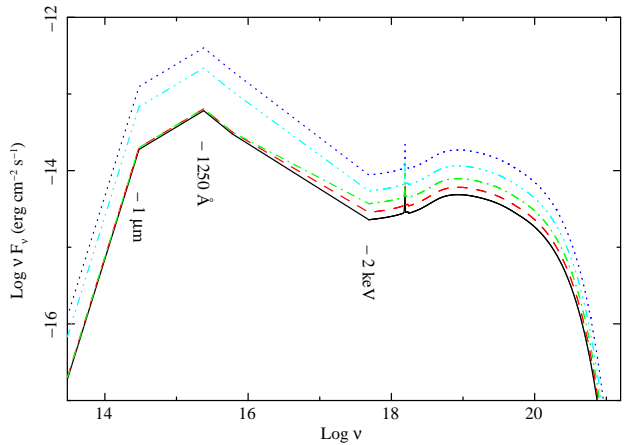}
 \caption{The model SEDs of the five sources: XBQ1 (black - solid line), XBQ2 (red - dashed line), XBQ3 (green - dot-dashed line), XBQ4 (dark blue - dotted line) and XBQ5 (light blue - triple dot-dashed line).}
\label{seds_plot}
\end{figure}

The model SEDs are plotted in Fig.~\ref{seds_plot}. They were used as the incident continuum models in version 2.1kn7 XSTAR \citep{kallman1982} calculations of the ionisation balance in the photoionised absorber, and the results of these model runs were collated into separate SPEX {\it xabs} model for each source. The SEDs were also used to create SPEX {\it file} table models for the source continua, to be used in fitting the X-ray spectra.

\subsection{Fitting the wind absorption}
\label{wind_fitting}

The X-ray spectral fitting was performed using SPEX 2.00.11. We fitted the combined EPIC X-ray spectrum of each source with models consisting of a renormalizable SED continuum (section~\ref{seds}) at the appropriate cosmological redshift, Galactic neutral absorption, and one ionised absorber phase for each of the BAL dynamical components apparent in the UV spectra. The outflow velocities and turbulent velocities of the absorber phases were fixed to the UV-derived values (Table~\ref{uv_abs_lines}); the `turbulent velocity' required by the SPEX \emph{xabs} model is the RMS velocity width, namely the measured FWHM of an absorption trough divided by 2.35. Where multiple UV absorbing components were present in a given source, they were each assumed to have equal absorbing columns $N_{H}$. The ionisation parameters of all velocity components in a given source were also constrained to be the same. 

There were thus three free non-tied parameters in each fit: continuum normalisation, ionisation parameter ($\xi$), and absorbing column ($N_{H}$). Although the X-ray continuum normalisation is in principle determined by the 2500~\AA\ luminosity, the X-ray and optical observations were not contemporaneous, and the sources are likely to be variable. The predicted 2~keV X-ray flux density is therefore only used in the construction of the SED, and the observed X-ray continuum normalisation was allowed to vary in the fitting. 

\begin{figure*}
\includegraphics[width=170mm,angle=-90]{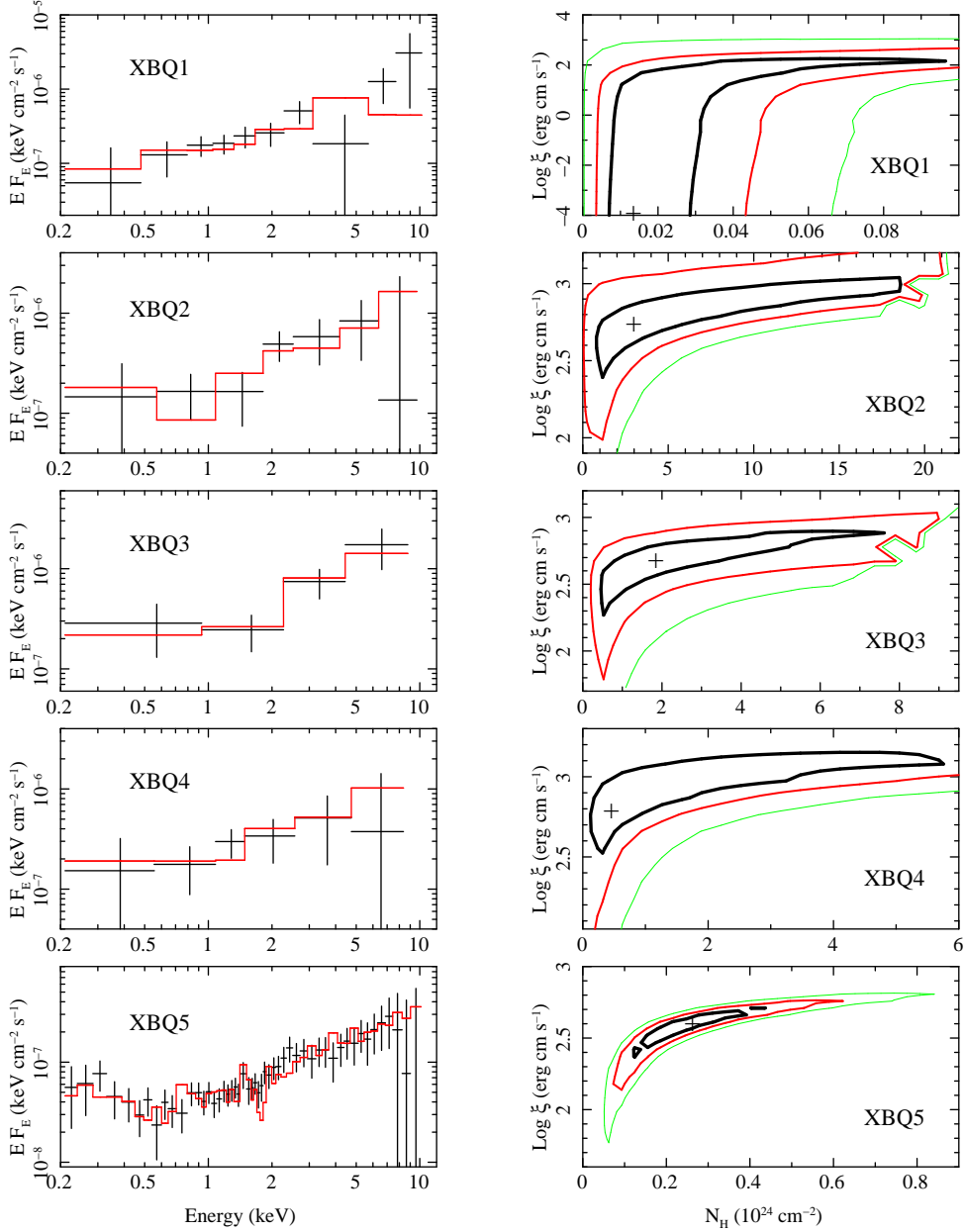}
 \caption{Left column: the EPIC combined X-ray spectra of the five XBALQSOs (observed frame) with the best-fit absorber models superimposed where appropriate. Right column: confidence contour plots of the free N$_{H}$ parameter versus log $\xi$; contours are 68.3\% (black), 95.4\% (red), and 99.7\% (green) confidence.}
\label{xray_fit_spectra}
\end{figure*}

The resulting fits are shown in Fig.~\ref{xray_fit_spectra}, alongside confidence contour plots for the free N$_{\rm H}$ and $\xi$ parameters. The best-fit values are listed in Table~\ref{xray_fit}. 

Interestingly, the fitted X-ray continuum normalisations are systematically higher than those predicted from the rest-frame UV flux; the ratios of fitted to predicted normalisations are listed in Table~\ref{xray_fit}. This corresponds with the finding in Section~\ref{fluxes_alpha_ox} that the median $\alpha_{ox}$ of the XBALQSOs is flatter than that of the OBALQSOs. To see whether the use of an X-ray richer SED would affect our results concerning the absorber properties, we generated a new SED for XBQ5 (for which we can obtain the best absorber constraints), multiplying the relative X-ray normalisation by the $N_{ratio}$ value for this source in Table~\ref{xray_fit}. We used this SED to generate a SPEX continuum model and an XSTAR ionised absorber model, and re-fitted the spectrum. We obtained almost identical values for $N_{H}$ and $\xi$ (well within 1\% of the original results), and so our assumption of a normal SED for the X-ray fitting should make no difference to our conclusions about the XBALQSO absorber properties.

\begin{table*}
 \centering
 \begin{minipage}{172mm}
  \caption{Absorption model fits to the X-ray spectra where the outflow speeds and turbulent velocities are held fixed at UV-derived values in each case: source name; $v_{shift}$ 1, velocity shift of dynamical phase 1 in km s$^{-1}$; $v_{turb}$ 1, RMS velocity width of dynamical phase 1 in km s$^{-1}$; $v_{shift}$ 2 and $v_{turb}$ 2 are the analogous quantities for dynamical phase 2; $N_{ratio}$, ratio of the fitted 1~keV continuum normalisation to that predicted from the 2500~\AA\, flux density using the \citet{vignali2003} relation; log $L_{x}$, log unabsorbed 2$-$10~keV (observed frame) X-ray luminosity in erg s$^{-1}$; log $\xi$, log ionisation parameter in erg cm s$^{-1}$; log $N_{H}$, log total equivalent Hydrogen column density of absorber in cm$^{-2}$ (i.e. twice the fitted column to a single velocity phase, where there are two phases); $\chi^{2}_{red}$, reduced chi-squared of the fits with degrees of freedom in brackets. }
\label{xray_fit}
  \begin{tabular}{@{}llllllllll@{}}
  \hline
Name  & $v_{shift}$ 1 & $v_{turb}$ 1 & $v_{shift}$ 2 & $v_{turb}$ 2 & $N_{ratio}$ & log $L_{x}$ & log $\xi$ & log $N_{H}$ & $\chi^{2}_{red}$ (d.o.f.) \\
  \hline
XBQ1 & -670   & 480  & -2170  & 1110 & 6 $\pm$ 1      & 44.08 $\pm$ 0.08   & -4$^{+6}_{-0}$\footnote{This is the low-ionization limit of the photoionized absorber model.} & 22.5 $\pm$ 0.2 & 1.53 (7) \\
XBQ2 & -1680  & 1030 & $-$    & $-$  & 185$^{+4\times10^{7}}_{-178}$  & 45$^{+5}_{-1}$      & 2.8$^{+0.2}_{-0.3}$    & 24.5$^{+0.3}_{-0.8}$ & 0.73 (4) \\
XBQ3 & -4100  & 1020 & -1330  & 620  & 360$^{+3\times10^{4}}_{-330}$ & 45$^{+2}_{-1}$ & 2.7$^{+0.1}_{-0.2}$ & 24.6$^{+0.4}_{-0.5}$ & 0.47 (1) \\
XBQ4 & -12740 & 1610 & -18750 & 2220 & 4$^{+210}_{-3}$ & 44.5$^{+1.7}_{-0.5}$ & 2.8$^{+0.2}_{-0.6}$    & 24.0$^{+0.4}_{-1.6}$ & 0.61 (3) \\
XBQ5 & -1900  & 830  & -4220  & 1370 & 8 $\pm$ 2      & 44.6 $\pm$ 0.1     & 2.62$^{+0.06}_{-0.14}$ & 23.7$^{+0.1}_{-0.2}$ & 0.87 (48) \\
\hline
\end{tabular}
\end{minipage}
\end{table*}

\section{The mass$-$energy budgets of the XBALQSOs and their X-ray absorbing winds}
\label{mass_energy_budget}

We used Eqn.~4 from \citet{warner2003} to calculate the black hole mass for each XBALQSO, with the predicted C~IV FWHMs (Table~\ref{lines}; see Section~\ref{dynamics}) and (dereddened) 1450~\AA\ luminosities obtained from our spectra; the errors are calculated according to the estimate of \citet{warner2003} that there is a 1$\sigma$ uncertainty on these masses of a factor of three. We then used the black hole masses to calculate the Eddington luminosity \citep[see~e.g.][]{krolik1999}. We estimated the bolometric luminosity from the monochromatic restframe 2500~\AA\, luminosity using the appropriate bolometric correction from \citet{richards2006}. These values are listed in Table~\ref{mass_energy}, along with the ratio $L_{bol}/L_{Edd}$, and the mass accretion rate which was estimated from $L_{bol}$ with an assumed accretion efficiency of 0.1. In estimating the error on $L_{bol}/L_{Edd}$, we assume that the uncertainty on $L_{bol}$ is far less than that on $L_{Edd}$; the error on $L_{Edd}$ is derived from the errors on the black hole masses. If the X-ray-to-optical flux ratio of the XBALQSOs is $\sim$~6 times higher than expected (see section~\ref{fluxes_alpha_ox}), then the bolometric correction we are using may underestimate $L_{bol}$ for these objects. For the \citet{marconi2004} SED, at a source luminosity of $10^{46}$~erg~s$^{-1}$, the $2-10$~keV X-ray band contains about 1.6\% of the bolometric luminosity; combining this with our figure for the flux ratio implies that $L_{bol}$ could be $\sim10\%$ higher than predicted with the standard bolometric correction.

It is interesting to estimate the mass outflow rates and kinetic luminosities of the winds, and to compare them with, respectively, the mass accretion rates and bolometric luminosities. For a radiatively-driven wind, it is possible to estimate these from the luminosity that is absorbed or scattered. Whether AGN winds are in fact radiatively driven has been the subject of much debate \citep[e.g.][]{crenshaw2000,krolik2001,chelouche2005}, but the winds must certainly be gaining some proportion of their momentum from the radiation which ionizes them. 

In the optically thin limit, a photon will generally only undergo at most one absorption or scattering event; the momentum of the wind will therefore be of the order of the momentum that it absorbs and scatters:

   \begin{equation}
   \dot M v = \frac{(L_{abs} + L_{scatt})}{c} \,,
\label{mom_approx}
   \end{equation}

where $\dot M$ is the mass outflow rate of the wind, $v$ is the outflow speed, $(L_{abs} + L_{scatt})$ is the luminosity absorbed or scattered, and $c$ is the speed of light. The quantity $(L_{abs} + L_{scatt})$ can be obtained straightforwardly from ionised absorber models (absorption modelled by SPEX \emph{xabs} components includes both photoelectric absorption and scattering with the Thompson and Klein-Nishina cross-sections), and so this allows a convenient estimation of $\dot M$ without the need to make assumptions about, for example, the distance of the wind from the ionising source. We have used this method in the past to calculate the mass outflow rate in nearby Seyferts, for whose absorbing columns (10$^{21}-10^{22}$~cm$^{-2}$) the optically thin approximation is reasonable \citep[e.g.][]{blustin2005}.

In BALQSOs, however, the much higher column absorbers are optically thick along our line of sight, and so photons encountering gas particles in the wind will scatter multiple times, giving up a far higher fraction of their kinetic energy than would be estimated via our momentum-transfer argument. In the limit of infinite optical depth, the wind's kinetic energy will be entirely derived from the radiative energy absorbed and scattered in it, i.e.

   \begin{equation}
   \frac{1}{2} \dot M v^{2} = (L_{abs} + L_{scatt}) \,.
\label{ke_approx}
   \end{equation}

We use Equations \ref{mom_approx} and \ref{ke_approx} to estimate, respectively, lower and upper limits to the mass outflow rates for the XBALQSOs. The resulting values for $\dot M$ can then be used to bracket the wind kinetic luminosity $L_{KE}$:

   \begin{equation}
      L_{KE} = \frac{1}{2} {\dot M} v^2  \,.
   \end{equation}

We list the derived values in Table~\ref{mass_energy}, along with the ratios $L_{KE}/L_{bol}$ and $\dot M/\dot M_{acc}$. 

\begin{table*}
 \centering
 \begin{minipage}{172mm}
  \caption{The mass-energy budgets of the XBALQSOS: Object; redshift; Log $M_{BH}$, log black hole mass in Solar masses; $\dot M_{acc}$, accretion rate in Solar masses per year; $\dot M$, total wind mass outflow rate in Solar masses per year; the ratio $\dot M/\dot M_{acc}$; Log $L_{bol}$, log bolometric luminosity in erg~s$^{-1}$; Log $L_{KE}$, total kinetic luminosity of wind in erg~s$^{-1}$; the ratio Log $L_{KE}/L_{bol}$; the ratio $L_{bol}/L_{Edd}$ where $L_{Edd}$ is the Eddington luminosity.}
\label{mass_energy}
  \begin{tabular}{@{}llllllllll@{}}
  \hline
Name & redshift & Log $M_{BH}$ & $\dot M_{acc}$ & $\dot M$ & $\dot M/\dot M_{acc}$ & Log $L_{bol}$ & Log $L_{KE}$ & Log $L_{KE}/L_{bol}$ & $L_{bol}/L_{Edd}$ \\
 \hline
XBQ1 & 2.63  & 8.2 $\pm$ 0.5 & 7.4 & 0.4$-$300 & 0.05$-$40   & 46.6 & 41$-$44 & -5$-$-3   & 2$^{+4}_{-1}$ \\
XBQ2 & 1.793 & 7.9 $\pm$ 0.5 & 3.2 & 8$-$3000  & 3$-$900     & 46.3 & 43$-$45 & -3$-$-0.8 & 1$^{+3}_{-1}$ \\
XBQ3 & 1.40  & 7.7 $\pm$ 0.5 & 1.1 & 20$-$7000 & 20$-$6000   & 45.8 & 44$-$46 & -2$-$0.05 & 0.9$^{+2}_{-0.6}$  \\
XBQ4 & 2.64  & 8.7 $\pm$ 0.5 & 49  & 0.2$-$7   & 0.004$-$0.1 & 47.4 & 43$-$45 & -4$-$-3   & 4$^{+7}_{-2}$ \\
XBQ5 & 2.82  & 8.1 $\pm$ 0.5 & 33  & 1$-$330   & 0.04$-$10   & 47.3 & 43$-$45 & -5$-$-2   & 10$^{+20}_{-7}$ \\
\hline
\end{tabular}
\end{minipage}
\end{table*}

\section{Discussion and conclusions}
\label{discussion}

Our goal in this paper was to investigate the X-ray and optical properties of a group of X-ray selected BALQSOs discovered in deep X-ray survey fields, and in particular to see whether they had been discoverable in such surveys due to having unusual spectral properties.

As we showed in section~\ref{fluxes_alpha_ox}, the XBALQSOs have, on average, slightly higher X-ray to optical flux ratios than the largest available comparable sample of OBALQSOs \citep{gallagher2006}. This is probably not a result of intrinsic source variability. It is equally possible that all five sources would happen to be especially X-ray faint as X-ray bright when we observed them, and if variability was a major factor, we would expect to see a mixture of high and low $\alpha_{ox}$ values among the XBALQSOs.

We can also immediately discard the possibility that the flatter $\alpha_{ox}$ is due to the XBALQSOs being intrinsically very luminous BALQSOs which happen to have high rest-frame optical extinction; as we showed in Section~\ref{extinction_properties}, their $B-Ks$ colours as a function of redshift are entirely consistent with the OBALQSO population. Given the very low space density of luminous QSOs on the sky, it is also highly unlikely that we should have found four such objects serendipitously in a single 30\arcmin\ diameter field.

\begin{figure}
\includegraphics[width=55mm,angle=-90]{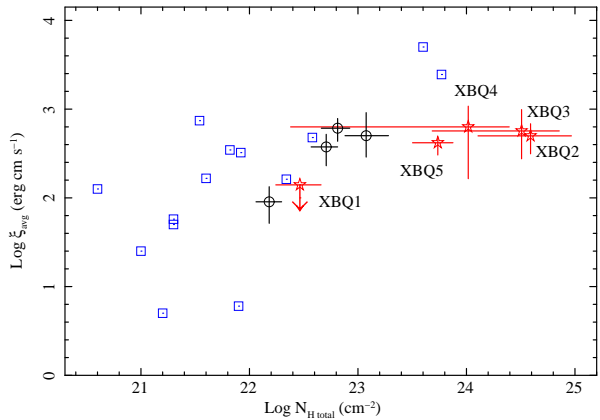}
 \caption{A comparison of total Log $N_{H}$ (the sum of $N_{H}$ for the individual phases) and column-weighted average log $\xi$ for the X-ray selected BALQSOs in our sample (red stars) with four optically-selected BALQSOs from the literature (black circles; see section~\ref{discussion}) and a sample of low$-$redshift `warm absorber' Seyferts and QSOs (blue squares; \citealt{blustin2005}).}
\label{xray_properties_avg}
\end{figure}

If XBALQSOs are brighter than expected in the X-rays due to having less intrinsic X-ray absorption than optically selected sources, the ionised X-ray absorbing winds should have lower N$_{H}$ and/or higher $\xi$ than those in comparable OBALQSOs. In Fig.~\ref{xray_properties_avg} we compare total absorbing column and (weighted average, where necessary) ionisation parameter with these quantities in OBALQSOs from the literature, as well as from a sample of lower redshift Seyfert galaxies and QSOs with ionised soft X-ray absorbers which we have studied previously \citep{blustin2005}. The OBALQSOs were PG1115+080 \citep{chartas2007}, Q1246-057, SBS 1542+541 \citep{grupe2003}, and PG2112+059 \citep{gallagher2004}. We find that, on the contrary, these XBALQSOs have similar or even perhaps greater intrinsic absorption than the OBALQSOs. Testing whether that is a general feature of XBALQSOs would require a larger sample of sources. 

The amount of absorption from the UV-absorbing part of the wind, as measured by the BI, is consistent with the distribution of BI for OBALQSOs in the SDSS, although there is again some indication that they are towards the more highly absorbed end of the distribution. 

The mass-energy transport of the winds, which we estimated in Section~\ref{mass_energy_budget}, ranges quite widely. For XBQ5, which has the best-constrained values, mass outflow rate via the X-ray absorbing part of the wind is $\sim$~0.04$-$10 times the mass accretion rate onto its black hole, and the kinetic luminosity of the wind is $\sim$~0.002$-$0.3\% of the bolometric luminosity. A complete picture of the mass-energy budget of the wind would require consideration of the mass-energy output via the UV-absorbing part of the wind, which is beyond the scope of this paper, and taking account of the full ionisation range of the wind, which would require far better data.

Regarding the nature of XBALQSOs, we are left with the conclusion that, although they are fundamentally similar to the wider population of OBALQSOs, they may belong to a part of the BALQSO population with an intrinsically higher X-ray to optical flux ratio. This is actually what we would expect for AGN as UV-faint as our XBALQSOs, since, according to the \citet{vignali2003} relation, for example, AGN with lower restframe 2500~\AA\ luminosities should have flatter $\alpha_{ox}$ indices. The implication of this is that, as hard X-ray surveys get deeper, there should be a lot of faint XBALQSO-type AGN waiting to be discovered. It is interesting to ask whether the XBALQSO $\alpha_{ox}$ values do in fact exactly follow the \citet{vignali2003} relation; our XBALQSOs do not span sufficient luminosity space for us to attempt an answer to that question. Investigating this point will require many more sources over a range of luminosities, with sufficient X-ray counts for a reliable estimation of the unabsorbed X-ray flux to be made.

\section*{Acknowledgments}

We wish to thank the anonymous referee for comments which improved the clarity of the paper. AJB acknowledges the support of an STFC Postdoctoral Fellowship. \emph{XMM-Newton} is an ESA science mission with instruments and contributions directly funded by ESA Member States and NASA. The Subaru Telescope is operated by the National Astronomical Observatory of Japan. The W.M. Keck Observatory is operated as a scientific partnership among the California Institute of Technology, the University of California and the National Aeronautics and Space Administration. The Observatory was made possible by the generous financial support of the W.M. Keck Foundation. The William Herschel Telescope is operated on the island of La Palma by the Isaac Newton Group in the Spanish Observatorio del Roque de los Muchachos of the Instituto de Astrofisica de Canarias. The Hale Telescope, Palomar Observatory, is part of a collaborative agreement between the California Institute of Technology, its divisions Caltech Optical Observatories and the Jet Propulsion Laboratory (operated for NASA), and Cornell University. The Two Micron All Sky Survey is a joint project of the University of Massachusetts and the Infrared Processing and Analysis Centre/California Institute of Technology, funded by the National Aeronautics and Space Administration and the National Science Foundation. The NASA/IPAC Infrared Science Archive is operated by the Jet Propulsion Laboratory, California Institute of Technology, under contract with the National Aeronautics and Space Administration. 

The authors wish to recognize and acknowledge the very significant cultural role and reverence that the summit of Mauna Kea has always had within the indigenous Hawaiian community.  We are most fortunate to have the opportunity to conduct observations from this mountain.

\label{lastpage}

\end{document}